\documentclass[10pt]{article}
\usepackage{amsmath}
\usepackage{amsfonts}
\usepackage{amssymb}
\usepackage{graphicx}
\usepackage[margin=2cm]{geometry}
\def\be{\begin{equation}}
\def\ee{\end{equation}}
\def\ba{\begin{align}}
\def\ea{\end{align}}
\title{On the smoothness of the multi-BMPV black hole spacetime}

\author{Graeme N. Candlish \\ \emph{University of Nottingham, University Park, Nottingham NG7 2RD, UK}}

\begin{document}

\maketitle

\begin{abstract}
We demonstrate that, in a multi-BMPV black hole spacetime, the event horizon is not smooth. We explicitly show that for a simpler configuration comprising a line of static extremal black holes and a single BMPV black hole, the metric at the horizon of the BMPV black hole is once, but not twice, continuously differentiable. We argue that this result is also valid when all the black holes are rotating. The Maxwell field strength is shown to be continuous, but not differentiable at the horizon. We also briefly demonstrate that previous work done to show lack of smoothness of static multi-centre solutions in five dimensions is not significantly modified by the inclusion of a higher derivative term in the action for five dimensional supergravity.
\end{abstract}

\section{Introduction}
Einstein-Maxwell theory admits multi-centre extremal black hole solutions, thanks to a balance of gravitational and electrostatic forces. Such a solution in four dimensions has been known for a long time, and the event horizon has been shown to be analytic \cite{hartlehawking}. In higher dimensions, however, such solutions do not generally have smooth event horizons \cite{Welch, Candlish}.

In \cite{Welch} this lack of smoothness was shown for a set of static black holes along a common axis by considering the behaviour of components of the Riemann tensor in a parallely propagated orthonormal frame along axial null geodesics. This result was extended in \cite{Candlish} by removing the restriction to axial geodesics and allowing an arbitrary number of co-axial static black holes to be present. Rather than working with the Riemann tensor, the behaviour of a geometric invariant was investigated, as such invariants will be as differentiable as the metric itself. In that work it was shown that the event horizon is generally $C^2$ but not $C^3$ in five dimensions\footnote{Five dimensional static multi-black holes on a Gibbons-Hawking base space were discussed in \cite{Kimura}, where the horizons were shown to be analytic. This is because the solutions are written in terms of harmonic functions on $\mathbb{R}^3$, and so the dependence of the radial coordinate on an affine parameter along a null geodesic is the same as that in four dimensions.}, and $C^1$ but not $C^2$ in more than five dimensions. Furthermore, in the case of more than five dimensions, it was demonstrated that there is a parallely propagated curvature singularity at the horizon. By tuning the parameters of the solution it is possible to increase the smoothness in five dimensions, and in fact it is possible to ensure that at least one connected component of the horizon is analytic. This is not the case for six dimensions or more.

In five dimensions, the action for Einstein-Maxwell-Chern-Simons theory coincides with the bosonic part of $\mathcal{N}=2$ five dimensional supergravity. All supersymmetric solutions to this theory have been classified according to whether there exists a timelike or null Killing vector \cite{allsolutions}. In the class of solutions corresponding to the timelike case, there are static black holes as discussed above, stationary rotating black holes \cite{breckenridge} (known as the BMPV solution), and black rings \cite{SUSYbr}.

As these solutions are supersymmetric (they satisfy a BPS bound), they are also extremal (the converse is not true), and thus multi-black hole solutions may be constructed due to the balance of forces exactly as described above. A striking example of such a solution comprising a set of concentric black rings is given in \cite{Gauntlett}. When the rings are coplanar the solution preserves both the $U(1)$ rotational isometries of a single supersymmetric black ring and the horizon is analytic.

We generally expect to see a non-analytic metric at the event horizon when a rotational symmetry is broken. A straightforward argument for this is given in \cite{horowitzreall}, where a supersymmetric black ring of non-constant charge density is shown not to admit a smooth horizon. The presence of a horizon for a rotating black hole requires a limit $\ell \to \infty$ where $\ell$ is the parameter along the direction of rotation. Any non-trivial periodic function of $\ell$ will therefore run through an infinite number of periods close to the horizon and so will not be continuous there. The behaviour of the parameter $\ell$ is familiar from the Kerr solution, where the azimuthal angle of the Boyer-Lindquist coordinates diverges at the horizon. It is essentially this effect we will see in this paper: a rotating black hole metric that has a broken symmetry in one of the directions of rotation will exhibit a lack of smoothness at the horizon.

One motivation of this work was to investigate the effects of a spin-spin interaction between two (or more) BMPV black holes. The lack of smoothness of the multi-BMPV spacetime could conceivably have been dependent on the relative alignments of the rotations, with parallel rotation resulting in a higher degree of differentiability than for the non-parallel case. A special case of parallel spins is when only one black hole rotates. Thus in this paper we consider an axial configuration of black holes, with a single, central rotating BMPV black hole, while the other black holes are static. The single BMPV solution has two planes of equal rotation and isometry group $U(1)^2 \times \mathbb{R}$. The axial configuration we are considering preserves only two of the isometries of a single BMPV black hole: time translations and a single $U(1)$ rotational symmetry. From the argument above we therefore expect lack of smoothness resulting from breaking the symmetry in a direction of rotation.

As we already know from \cite{Candlish} that the five dimensional extremal static multi-black hole solution only admits a $C^2$ horizon, we might ask if our special case of a single rotating black hole amongst static black holes is at least as smooth. Somewhat surprisingly we find that this is not the case, and we will demonstrate that our configuration exhibits an event horizon that is $C^1$ but not $C^2$. Thus we can conclude that a general multi-BMPV solution must not have a $C^2$ horizon.

At the end of this paper we will return briefly to the purely static case, this time in the context of higher derivative theories. Following closely the work of \cite{higherderiv} we will show that the degree of differentiability of a multi-centre static solution is not affected by the inclusion of a four derivative term in the supergravity action.

To begin with then, in section \ref{first} we briefly describe some important aspects of the single BMPV black hole solution, as well as the Gaussian null coordinates which are central to our work. In section \ref{second} we give the metric for the axial BMPV/static black hole spacetime we are considering. The main analysis and result is discussed in section \ref{fourth}. The discussion of higher derivative corrections to static multi-centre solutions is given in section \ref{fifth}. Finally we conclude and summarise the results.

\section{Single BMPV Solution}
\label{first}
The BMPV black hole solution to minimal five dimensional supergravity has the following metric:
\be
\label{bps}
ds^2 = -H^{-2} (dt + \omega)^2 + H ds^2(\mathbb{R}^4).
\ee
The function $H$ and the one-form $\omega$ are defined on $\mathbb{R}^4$. $H$ satisfies the four dimensional Laplace equation and is therefore harmonic. Poles in $H$ correspond to the location of the event horizon of a black hole. Thus for a single black hole at the origin of a Cartesian coordinate system we have
\be
H = 1 + \frac{\mu}{|x|^2}
\ee
where $\mu$ is a parameter proportional to the mass. As we have four spatial dimensions there are two independent planes of rotation. Regularity of the horizon demands that the BMPV black hole has equal rotations in both two-planes. Thus the four dimensional rotation group $SO(4) \cong SU(2) \times SU(2)$ is effectively restricted to one of the $SU(2)$ factors. The one-form $\omega$ describes the rotation of the black hole, and is given by
\be
\label{omega}
\omega = \frac{J}{2\mu} K_{ij} \partial_i H dx^j = \frac{J}{2|x|^4} K_{ji} x^i dx^j
\ee
where $J$ is a parameter proportional to the angular momentum and $K_{ji}$ are the components of a self-dual two-form which describes rotation in two orthogonal two-planes of $\mathbb{R}^4$. An arbitrary rotation is described by a linear combination of three such two-forms $K^{\alpha}$ (with $\alpha = 1,2,3$) which, as matrices on $\mathbb{R}^4$, satisfy the commutation relations of the $su(2)$ algebra:
\be
\label{su2}
[K^{\alpha},K^{\beta}] = 2\epsilon_{\alpha \beta \gamma} K^{\gamma}.
\ee
For a single black hole we may, without loss of generality, choose the planes of rotation to be in the $x^1,x^2$ and $x^3,x^4$ directions. Thus the rotation one-form may be written
\be
\omega = \frac{J}{2|x|^4} (x^1 dx^2 - x^2 dx^1 + x^3 dx^4 - x^4 dx^3).
\ee
Working in a polar coordinate system specified by
\begin{equation}
\begin{split}
\label{coordinates}
x^1 &= R \sin \theta \cos \phi \\
x^2 &= R \sin \theta \sin \phi \\
x^3 &= R \cos \theta \cos \psi \\
x^4 &= R \cos \theta \sin \psi
\end{split}
\end{equation}
where $0 \; \leq \; \theta \; \leq \; \pi$ and $0 \; \leq \; \phi,\psi \; < \; 2\pi$, we can write the metric as
\be
\label{mainmetric}
ds^2 = -H^{-2} (dt + \omega)^2 + H (dR^2 + R^2(d\theta^2 + \sin^2 \theta d\phi^2 + \cos^2 \theta d\psi^2))
\ee
with
\be
\label{singleBMPV}
H = 1 + \frac{\mu}{R^2}, \quad \omega = \frac{J}{2R^2} (\sin^2 \theta d\phi + \cos^2 \theta d\psi).
\ee
The event horizon is located at $R=0$ and we have a coordinate singularity there. We will transform to a Gaussian null coordinate system to ensure a regular metric on the horizon. It is clear from this that the single BMPV black hole has three isometries: time translations and two $U(1)$ rotational symmetries parameterised by $\phi$ and $\psi$.

\subsection{Gaussian null coordinates}
To investigate the smoothness of the multi-BMPV solution we will put the metric into a Gaussian null system of coordinates. The method of construction of such a coordinate system is discussed in \cite{reallsusy} and in detail in \cite{Wald}. Here we present only a brief description.

Consider a connected component of the future event horizon $\mathcal{H}^+$ of our multi-black hole spacetime, which we will call $\mathcal{H}^+_0$. Let us label the intersection of this component of the horizon with a spatial hypersurface as $H_0$. The surface $H_0$ is topologically an $S^3$, on which we introduce the coordinates $x^i$. In the coordinates in (\ref{mainmetric}) the time translation Killing vector is given by $V = \partial/\partial_t$. As this is null on the horizon and the generator of a symmetry, this is tangent to the null geodesic generators of $\mathcal{H}_0^+$. We now define $v$ to be the parameter-distance from $H_0$ along the integral curves of $V$. This gives us a coordinate chart $\{v,x^i\}$ on a neighbourhood of $H_0$ in $\mathcal{H}_0^+$.

We now extend this chart off the horizon along a null geodesic. Let $U$ be the unique (past-directed) null vector field satisfying $U \cdot V = 1$ and $U \cdot \partial/\partial x^i = 0$ on $\mathcal{H}_0^+$. We take $\gamma(v,x^i)$ to be the null geodesic that begins at the point with coordinates $\{v,x^i\}$ in $\mathcal{H}_0^+$ and whose tangent vector there is $U$. The affine parameter distance from $\mathcal{H}_0^+$ along this geodesic is given by $\lambda$. The Gaussian null coordinates on a neighbourhood of $H_0$ are therefore $\{v,\lambda,x^i\}$, and the metric takes the general form
\begin{equation}
ds^2 = -H(\lambda,x)^{-2}dv^2 + 2dvd\lambda + 2\lambda h_i(\lambda,x)dvdx^i + h_{ij}(\lambda,x)dx^i dx^j.
\end{equation}

To illustrate this we will now transform the metric given by (\ref{mainmetric}) and (\ref{singleBMPV}) to Gaussian null coordinates. From time translation symmetry, and the choice of a past-directed geodesic, we have
\begin{equation}
\label{tgeo}
-H^{-2}(\dot{t} + \dot{\omega}) = E,
\end{equation}
where the dot denotes differentiation with respect to the affine parameter $\lambda$, and $\dot{\omega} = \omega_{\phi} \dot{\phi} + \omega_{\psi} \dot{\psi}$. We choose the parameterisation such that we can set the constant of the motion $E=1$. The rotational symmetries parametrised by $\phi$ and $\psi$ give us
\begin{equation}
\begin{split}
\label{anglegeo}
\dot{\phi} &= \frac{J_{\phi} - \omega_{\phi}}{HR^2\sin^2 \theta} \\
\dot{\psi} &= \frac{J_{\psi} - \omega_{\psi}}{HR^2\cos^2 \theta}
\end{split}
\end{equation}
where $J_{\phi}$ and $J_{\psi}$ are the angular momenta of the geodesic in the $\phi$ and $\psi$ direction respectively. By definition of the Gaussian null coordinate system, we require the tangent vector to the geodesic to be orthogonal to the vector fields tangent to the $S^3$ of $H_0$. Therefore we set
\begin{equation}
\label{initcond}
\dot{\theta}|_{\mathcal{H}^+} = J_{\phi}|_{\mathcal{H}^+} = J_{\psi}|_{\mathcal{H}^+} = 0.
\end{equation}
Note that we choose a geodesic where $\theta$ is constant. Given this we now have
\begin{equation}
\label{angleeqn}
\dot{\phi} = \dot{\psi} = -\frac{J}{2HR^4}.
\end{equation}
Substituting (\ref{tgeo}) and (\ref{anglegeo}) into the null condition, using (\ref{initcond}) and the expression for $\omega$ in (\ref{singleBMPV}), we find
\begin{equation}
\label{Reqn}
\dot{R} = \left( H - \frac{J^2}{4H^2R^6} \right)^{1/2}
\end{equation}
This can be expanded out for small $R$, integrated and the resulting series inverted to obtain
\begin{equation}
\label{Rlambda}
R(\lambda) = \left( \frac{\Delta}{\mu} \right)^{1/2} \lambda^{1/2} \left(1 + \frac{J^2 + 2\mu^3}{4\mu^2\Delta}\lambda + \mathcal{O}(\lambda^2) \right)
\end{equation}
where the term in brackets is analytic in $\lambda$ and we have defined the symbol
\begin{equation}
\Delta \equiv (4\mu_1^3 - J^2)^{1/2}.
\end{equation}
To ensure the absence of closed timelike curves outside the horizon, as discussed in \cite{Gibbons}, we must have $\Delta > 0$. Clearly this is an outgoing null geodesic for increasing $\lambda$. Given this and (\ref{anglegeo}), along with the initial conditions on our geodesic we can find
\begin{equation}
\begin{split}
\label{anglelambda}
\phi &= -\frac{J}{2\Delta}\ln \lambda + \Phi + \frac{J(10\mu^3 - J^2)}{4\Delta^2 \mu^2}\lambda + \mathcal{O}(\lambda^2) \\
\psi &= -\frac{J}{2\Delta}\ln \lambda + \Psi + \frac{J(10\mu^3 - J^2)}{4\Delta^2 \mu^2}\lambda + \mathcal{O}(\lambda^2)
\end{split}
\end{equation}
with $\Psi$ and $\Phi$ constants of integration. We use these as two of the coordinates on the $S^3$ as they are finite and 
\be
\partial/\partial_{\phi} = \partial/\partial_{\Phi}, \quad \partial/\partial_{\psi} = \partial/\partial_{\Psi}.
\ee
The final coordinate on the $S^3$ is given by $\Theta$ which is the limiting value of $\theta$ along the geodesic at the horizon. In this case we are taking $\theta$ to be constant along the geodesic, so $\theta = \Theta$. This will not be true for the multi-black hole spacetime. As discussed earlier, the transformation to coordinates regular on the horizon requires an infinite shift at the horizon in coordinates parameterising directions of rotation. Any dependence in the metric on these coordinates will lead to a non-analytic horizon, as we will see later. The coordinates on the $S^3$, previously denoted $x^i$, are now given by $\{\Theta,\Phi,\Psi\}$. It is now straightforward to determine the required transformations to put the metric in Gaussian null coordinates. Using (\ref{Rlambda}) to rewrite $H$ as a function of $\lambda$, we can integrate (\ref{tgeo}) to get
\begin{equation}
v \equiv t + \int (H^2 + \dot{\omega})d\lambda.
\end{equation}
Thus we transform the $t$ coordinate according to
\begin{equation}
dt = dv - \left( H(\lambda)^2 - \frac{J^2}{4H(\lambda)R(\lambda)^6} \right) d\lambda.
\end{equation}
Likewise, using (\ref{Reqn}) and (\ref{angleeqn}), the transformations for the $R$, $\phi$ and $\psi$ coordinates are given by
\begin{equation}
\begin{split}
dR &= \left( H(\lambda) - \frac{J^2}{4H(\lambda)^2R(\lambda)^6} \right)^{1/2} d\lambda \\
d\phi &= d\Phi - \frac{J}{2H(\lambda)R(\lambda)^4} d\lambda \\
d\psi &= d\Psi - \frac{J}{2H(\lambda)R(\lambda)^4} d\lambda,
\end{split}
\end{equation}
and $d\theta = d\Theta$. The metric in Gaussian null coordinates is
\begin{multline}
ds^2 = -H(\lambda)^{-2}dv^2 + 2dvd\lambda - \frac{J}{H(\lambda)^2R(\lambda)^2} dv \left( \sin^2 \theta d\Phi + \cos^2 \theta d\Psi \right) + H(\lambda)R(\lambda)^2 d\Theta^2 \\ + \left( H(\lambda)R(\lambda)^2 \sin^2 \Theta - \frac{J^2 \sin^4 \Theta}{4H(\lambda)^2R(\lambda)^4} \right)d\Phi^2 + \left( H(\lambda)R(\lambda)^2 \cos^2 \Theta - \frac{J^2 \cos^4 \Theta}{4H(\lambda)^2R(\lambda)^4} \right)d\Psi^2 \\ - \frac{J^2 \sin^2 (2\Theta)}{8H(\lambda)^2R(\lambda)^4} d\Phi d\Psi.
\end{multline}
The transformation of the multi-black hole spacetime to a Gaussian null coordinate system will proceed in the same way.

\section{Axial Multi-Black Hole Spacetime}
\label{second}
We now give the metric for the configuration of black holes in which we are interested: a single BMPV black hole in a line of extremal static black holes. As argued in the introduction, the result of the smoothness analysis for this special case will apply to a general multi-BMPV spacetime. The metric in this case is again given by
\begin{equation}
\label{multimetric}
ds^2 = -H^{-2}(dt + \omega)^2 + H (dR^2 + R^2(d\theta^2 + \sin^2 \theta d\phi^2 + \cos^2 \theta d\psi^2))
\end{equation}
but the harmonic function is now
\begin{equation}
\label{multiharm}
H = 1 + \frac{\mu_1}{R^2} + \sum_{A=2}^N \frac{\mu_A}{R^2 + a_A^2 - 2a_A R \sin \theta \sin \phi}
\end{equation}
where $\mu_A$ and $a_A$ denote, respectively, the mass parameter and position of the $A$-th black hole. The black holes are arrayed along the $x^2$ axis in Cartesian coordinates. We can rewrite this as
\begin{equation}
\label{multiharmgegen}
H = \frac{\mu_1}{R^2} + \sum_{n=0}^{\infty} h_n R^n Y_n (\sin \theta \sin \phi)
\end{equation}
where the harmonics $Y_n(\sin \theta \sin \phi)$ are given by the Gegenbauer polynomials
\begin{equation}
Y_n(\sin \theta \sin \phi) = C^1_n (\sin \theta \sin \phi).
\end{equation}
The coefficients are
\begin{equation}
\label{hn}
h_n = \delta_{n,0} + \sum_{A=2}^N \frac{\mu_A}{a_A^{2+n}}.
\end{equation}
The rotation one-form is as before for a single BMPV black hole:
\begin{equation}
\omega = \frac{J}{2R^2} (\sin^2 \theta d\phi + \cos^2 \theta d\psi )
\end{equation}
and we see that now we have two Killing symmetries: time translations and one of the rotational $U(1)$ isometries, parameterised by $\psi$.

\section{Smoothness Analysis}
\label{fourth}
The smoothness of the multi-black hole metric will be determined by examining the behaviour of a geometric invariant of the solution along a null geodesic. The invariant we consider is the norm of the Killing vector field $K = \partial/\partial_{\psi}$. As argued in \cite{Candlish}, the Killing vector fields will have the same degree of differentiability as the full metric. To be more precise, we will determine $K^2$ as an expansion in the affine parameter $\lambda$ close to the horizon\footnote{That is, the connected component of $\mathcal{H}^+$ corresponding to the BMPV black hole.} at $\lambda = 0$. We will see that this expansion contains terms that are not twice continuously differentiable at the horizon. Using the null geodesic we will then transform the metric given by (\ref{multimetric}) and (\ref{multiharm}) into a Gaussian null coordinate system to determine the precise degree of differentiability.

\subsection{Geodesic equations}
First we must solve the geodesic equations to determine a past-directed outgoing null geodesic. As before we have
\begin{equation}
\dot{t} + \dot{\omega} = -H^{2}
\end{equation}
where $\dot{\omega} = \omega_{\phi} \dot{\phi} + \omega_{\psi} \dot{\psi}$, and again we have chosen the parametrisation such that the constant of the motion $E=1$. We integrate to get
\begin{equation}
\label{vdef}
v \equiv t + \int (H^{2} + \dot{\omega}) d\lambda.
\end{equation}
The remaining rotational symmetry again gives
\begin{equation}
\label{psigeo}
\dot{\psi} = \frac{J_{\psi} - \omega_{\psi}}{HR^2 \cos^2 \theta}.
\end{equation}

The geodesic equation (after substituting for $\dot{t}$) for $R$ is
\begin{multline}
\ddot{R} - R \dot{\theta}^2 - R \dot{\phi}^2 \sin^2 \theta - R \dot{\psi}^2 \cos^2 \theta + H^{-1} \dot{R} (\dot{\phi} \partial_{\phi} H + \dot{\theta} \partial_{\theta} H) \\ + H^{-1} (\dot{\psi} f_{\psi R} + \dot{\phi} f_{\phi R}) - \partial_R H + \frac{1}{2} H^{-1} \dot{R}^2 \partial_R H \\ - \frac{1}{2} H^{-1} R^2 \partial_R H (\dot{\theta}^2 + \dot{\psi}^2 \cos^2 \theta + \dot{\phi}^2 \sin^2 \theta) = 0.
\end{multline}
The geodesic equation for $\theta$ is
\begin{multline}
\label{thetageo}
\ddot{\theta} + 2 R^{-1} \dot{R} \dot{\theta} + \frac{1}{2} \dot{\psi}^2 \sin (2\theta) - \frac{1}{2} \dot{\phi}^2 \sin (2\theta) + H^{-1} \dot{\theta} (\dot{\phi} \partial_{\phi} H + \dot{R} \partial_R H) \\ + H^{-1} R^{-2} (\dot{\psi} f_{\psi \theta} + \dot{\phi} f_{\phi \theta}) - R^{-2} \partial_{\theta} H + \frac{1}{2} H^{-1} \dot{\theta}^2 \partial_{\theta} H \\ - \frac{1}{2} H^{-1} \partial_{\theta} H (R^{-2} \dot{R}^2 + \dot{\psi}^2 \cos^2 \theta + \dot{\phi}^2 \sin^2 \theta) = 0.
\end{multline}
The geodesic equation for $\phi$ is
\begin{multline}
\ddot{\phi} + 2 R^{-1} \dot{R} \dot{\phi} + 2 \cot \theta \dot{\theta} \dot{\phi} - R^{-2} \csc^2 \theta \partial_{\phi} H \\ + \frac{1}{2} H^{-1} \csc^2 \theta \partial_{\phi} H \left( \sin^2 \theta \dot{\phi}^2 - \cos^2 \theta \dot{\psi}^2 - R^{-2} \dot{R}^2 - \dot{\theta} \right) \\ - R^{-2} H^{-1} \csc^2 \theta \left( \dot{R} f_{\phi R} + \dot{\theta} f_{\phi \theta} \right) + H^{-1}\dot{\phi} \left( \dot{R} \partial_R H + \dot{\theta} \partial_{\theta} H \right) = 0.
\end{multline}
The final equation we require is the null condition,
\begin{equation}
-H^2 + H(\dot{R}^2 + R^2 \dot{\theta}^2 + R^2 \dot{\phi}^2 \sin^2 \theta + R^2 \dot{\psi}^2 \cos^2 \theta)=0.
\end{equation}
We will solve these equations using a power series in the affine parameter $\lambda$.

\subsection{Solving the geodesic equations}
\label{justify}
Before we give the expressions for the power series ansatz we will use to solve the geodesic equations, we will first give an idea of what we expect them to look like. The geodesic equation for $\theta$ turns out to decouple from the others, at least up to the order relevant for our current purpose. Therefore, we will calculate the first non-trivial terms in the $\theta$ expansion directly. We are aided in this task by the fact that our solution should look like that for a single BMPV black hole very near the horizon. Thus we can immediately write, given (\ref{anglelambda}), the following form for the angles as functions of $\lambda$:
\begin{equation}
\begin{split}
\phi &= \Phi + w_0 \ln \lambda + w_2 \lambda + \delta \phi \\
\psi &= \Psi + y_0 \ln \lambda + y_2 \lambda + \delta \psi \\
\end{split}
\end{equation}
where the corrections $\delta \phi$ and $\delta \psi$ are assumed to be terms of higher order than $\mathcal{O}(\lambda)$. Similarly for $R$ we have
\begin{equation}
R = a_1 \lambda^{1/2} + a_3 \lambda^{3/2} + \delta R
\end{equation}
with $\delta R$ being a term of higher order than $\mathcal{O}(\lambda^{3/2})$. We merely assume, again using the behaviour found earlier for a single BMPV black hole, that the leading order term of $\theta$ is a constant, i.e. $\theta = \Theta + \delta \theta$. We substitute these expressions into (\ref{thetageo}) and only keep the terms of lowest order in $\lambda$. We expand out functions such as $\sin(\phi)$ for small $\lambda$ using normal trigonometric identities and performing a Taylor expansion on coefficients that do not involve any $\ln \lambda$ terms. The first non-trivial terms in the geodesic equation appear at order $\lambda^{-1/2}$, leading to the following second order ordinary differential equation:
\begin{equation}
\ddot{\theta} - \left(2 \frac{h_1}{a_1} + \frac{1}{4\mu_1} a_1^3 h_1 + \frac{a_1^3}{\mu_1} h_1 w_0^2 \right) \cos \Theta \left( \cos \Phi \sin (w_0 \ln \lambda) + \sin \Phi \cos (w_0 \ln \lambda) \right)\lambda^{-1/2} = 0.
\end{equation}
This is solved by
\begin{equation}
\theta(\lambda) = \Theta + b_2 \lambda + \frac{4c \left(4 \sin (\Phi + w_0 \ln \lambda) w_0^2 - 3 \sin (\Phi + w_0 \ln \lambda) + 8 \cos (\Phi + w_0 \ln \lambda)w_0 \right) \lambda^{3/2} }{9 + 40w_0^2 + 16w_0^4},
\end{equation}
with $b_2$ a constant and $c$ given by
\begin{equation}
c = \left( 2 \frac{h_1}{a_1} + \frac{a_1^3}{\mu_1} h_1 \left( \frac{1}{4} + w_0^2 \right) \right) \cos \Theta.
\end{equation}
Thus we see that terms such as $\sin(m w_0 \ln \lambda)$ and $\cos(m w_0 \ln \lambda)$, with positive integer $m$, arise in the expansion of $\theta$. Likewise, such terms will arise at the same order in $\phi$ and $\psi$, and at order $\lambda^2$ in $R$.

\subsection{Series ansatz}
\label{seriesansatz}
Given the form found above, and using the known leading order behaviour of null geodesics for a single BMPV black hole that we found earlier, we use the following series expansions of $R, \theta, \phi$ and $\psi$. For $R$ we have
\begin{align}
R &= a_1 \lambda^{1/2} + a_3 \lambda^{3/2} + \sum_{n=5}^{\infty} a_n(\Theta,\Phi) \lambda^{n/2} \\ &+ \sum_{k=4}^{\infty} \lambda^{k/2} \left( \sum_{m=1}^{k-3} a_k^{(m)}(\Theta,\Phi) \sin (m w_0 \ln \lambda) + \tilde{a}_k^{(m)}(\Theta,\Phi) \cos (m w_0 \ln \lambda) \right). \\
\end{align}
After solving the geodesic equations we have found that $a_n = 0$ if $n$ is an even integer, and $a_k^{(m)} = \tilde{a}_k^{(m)} = 0$ if $k$ and $m$ are both even or both odd.

For the coordinates $\phi$ and $\psi$, the leading term in the ansatz is the logarithmic behaviour we found earlier for the single BMPV black hole, while the leading order term in $\theta$ is a constant. Thus we have
\begin{align}
\theta &= \Theta + b_2 \lambda + \sum_{N=4}^{\infty} b_N(\Theta,\Phi) \lambda^{N/2} \\ &+ \sum_{k=3}^{\infty} \lambda^{k/2} \left( \sum_{M=1}^{k-2} b_k^{(M)}(\Theta,\Phi) \sin (M w_0 \ln \lambda) + \tilde{b}_k^{(M)}(\Theta,\Phi) \cos (M w_0 \ln \lambda) \right) \\
\phi &= \Phi + w_0 \ln \lambda + w_2 \lambda + \sum_{N=4}^{\infty} w_N(\Theta,\Phi) \lambda^{N/2} \\ &+ \sum_{k=3}^{\infty} \lambda^{k/2} \left( \sum_{M=1}^{k-2} w_k^{(M)}(\Theta,\Phi) \sin (M w_0 \ln \lambda) + \tilde{w}_k^{(M)}(\Theta,\Phi) \cos (M w_0 \ln \lambda) \right) \\
\psi &= \Psi + y_0 \ln \lambda + y_2 \lambda + \sum_{N=4}^{\infty} y_N(\Theta,\Phi) \lambda^{N/2} \\ &+ \sum_{k=3}^{\infty} \lambda^{k/2} \left( \sum_{M=1}^{k-2} y_k^{(M)}(\Theta,\Phi) \sin (M w_0 \ln \lambda) + \tilde{y}_k^{(M)}(\Theta,\Phi) \cos (M w_0 \ln \lambda) \right).
\end{align}
Again, after solving the geodesic equations, we have that $b_N = w_N = y_N = 0$ if $N$ is an odd integer, and $b_k^{(M)} = \tilde{b}_k^{(M)} = w_k^{(M)} = \tilde{w}_k^{(M)} = y_k^{(M)} = \tilde{y}_k^{(M)} = 0$ if $M$ is odd and $k$ is even or vice versa.

It turns out that we only need terms up to order $\lambda^2$ in $R(\lambda)$ and order $\lambda^{3/2}$ in the angles to determine the degree of differentiability of the metric. All these coefficients will, in general, depend on the parameters of the solution, that is
\begin{equation}
a_i = a_i (\mu_1,J,h_n,\Phi,\Theta)
\end{equation}
and likewise for $b_i$, $y_i$ and $w_i$.

\subsection{Coefficients}
Using computer algebra, we expand each geodesic equation as a series in $\lambda$. Then, for each order in $\lambda$, we solve for the coefficients in the ansatz.

Therefore we find $y_0$ in terms of $a_1$ by solving the $\psi$ geodesic equation at order $\lambda^{-2}$ (leading order). Similarly we can find $w_0$ by solving the $\phi$ equation at the same order to find
\begin{equation}
w_0 = y_0 = -\frac{J}{2\Delta}.
\end{equation}
as expected from the single BMPV case. The $\theta$ equation has non-zero terms at order $\lambda^{-2}$, but the equation has an overall factor of $(w_0 - y_0)$ which vanishes as $w_0 = y_0$. The $R$ equation at leading order ($\mathcal{O}(\lambda^{-3/2})$) leads to a quartic polynomial in $a_1$, and so there are 4 solutions. Two are complex, and of the two real solutions we take the positive one:
\begin{equation}
a_1 = \left( \frac{\Delta}{\mu_1} \right)^{1/2}
\end{equation}
which agrees with that found earlier for the single BMPV black hole, as expected. The coefficients $w_2,y_2$ and $b_2$ are unfixed by the equations, being set by the initial conditions on the geodesic. These conditions are again given by (\ref{initcond}). In terms of the coefficients this means
\begin{equation}
\begin{split}
b_2 &= 0 \\
w_2 &= y_2 = \frac{J h_0 \left(10 \mu_1^3 - J^2 \right)}{4 \Delta^2 \mu_1^2}.
\end{split}
\end{equation}
Comparing this with (\ref{anglelambda}) we see this agrees with the $\mathcal{O}(\lambda)$ terms there, upon setting $h_0 = 1$ for a single black hole. The next few non-vanishing coefficients in the expansion of $R$ are
\begin{equation}
\begin{split}
a_3 &= \frac{h_0 \left(2 \mu_1^3+J^2\right)}{4 \mu_1^{5/2} \sqrt{\Delta}} \\
a_4^{(1)}(\Theta,\Phi) &= - \frac{\sin (\Theta ) h_1 \Delta \left(2 J \left(J \cos (\Phi )+2 \sin (\Phi ) \Delta \right)-5 \cos (\Phi ) \mu_1^3\right)}{25 \mu_1^6-6 J^2 \mu_1^3} \\
\tilde{a}_4^{(1)}(\Theta,\Phi) &= - \frac{\sin (\Theta ) h_1 \Delta \left(2 J \left(J \sin (\Phi )-2 \cos (\Phi ) \Delta \right)-5 \sin (\Phi ) \mu_1^3\right)}{25 \mu_1^6-6 J^2 \mu_1^3}.
\end{split}
\end{equation}
The next non-trivial coefficients in the expansion of $\theta$ are
\begin{equation}
\begin{split}
b_3^{(1)}(\Theta,\Phi) &= -\frac{3 \cos (\Theta ) h_1 \Delta^{3/2} \left(J \left(J \cos (\Phi )+\sin (\Phi ) \Delta \right)-3 \cos(\Phi ) \mu_1^3\right)}{\mu_1^{5/2} \left(9 \mu_1^3-2 J^2\right)} \\
\tilde{b}_3^{(1)}(\Theta,\Phi) &= \frac{3 \cos (\Theta ) h_1 \Delta^{3/2} \left(3 \sin (\Phi ) \mu_1^3+J \left(\cos (\Phi ) \Delta -J \sin(\Phi )\right)\right)}{\mu_1^{5/2} \left(9 \mu_1^3-2J^2\right)}.
\end{split}
\end{equation}
For the next order in the $\phi$ expansion we have
\begin{equation}
\begin{split}
w_3^{(1)}(\Theta,\Phi) &= \left[ \csc (\Theta ) h_1 \left(4 \mu_1^3-J^2\right){}^{3/4} \left(-450 \sin (\Phi ) \mu_1^6+3 J \left(J (13 \cos (2 \Theta )+73) \sin (\Phi ) \right. \right. \right. \\ &\left. \left. \left. -5 (3 \cos (2 \Theta )+7) \cos (\Phi ) \Delta \right) \mu_1^3+2 J^3 \left((7 \cos (2 \Theta )+11) \cos (\Phi ) \Delta \right. \right. \right. \\ & \left. \left. \left. -J (5 \cos (2 \Theta )+13) \sin (\Phi )\right)\right)\right] / \left[ 2 \mu_1^{5/2} \left(225 \mu_1^6-104 J^2 \mu_1^3+12J^4\right) \right] \\
\tilde{w}_3^{(1)}(\Theta,\Phi) &= \left[ \csc (\Theta ) h_1 \Delta^{3/2} \left(450 \cos (\Phi ) \mu_1^6-3 J \left(J (13 \cos (2 \Theta )+73) \cos (\Phi ) \right. \right. \right. \\ &\left. \left. \left. +5 (3 \cos (2 \Theta )+7) \sin (\Phi ) \Delta \right) \mu_1^3+2 J^3 \left(J (5 \cos (2 \Theta )+13) \cos (\Phi ) \right. \right. \right. \\ &\left. \left. \left. +(7 \cos (2 \Theta )+11) \sin(\Phi ) \Delta \right)\right)\right] / \left[ 2 \mu_1^{5/2} \left(225 \mu_1^6-104 J^2 \mu_1^3+12J^4\right) \right].
\end{split}
\end{equation}
At this order the coefficients for $\psi$ are
\begin{equation}
\begin{split}
y_3^{(1)}(\Theta,\Phi) &= -\left[ J \sin (\Theta ) h_1 \Delta^{3/2} \left(\left(39 J \sin (\Phi )-45 \cos (\Phi ) \Delta \right) \mu_1^3 \right. \right. \\ &\left. \left. +2 J^2 \left(7 \cos (\Phi ) \Delta -5 J \sin (\Phi )\right)\right) \right] / \left[ \mu_1^{5/2} \left(225 \mu_1^6-104 J^2 \mu_1^3+12 J^4\right) \right] \\
\tilde{y}_3^{(1)}(\Theta,\Phi) &= -\left[ J \sin (\Theta ) h_1 \Delta^{3/2} \left(2 J^2 \left(5 J \cos(\Phi )+7 \sin (\Phi ) \Delta \right) \right. \right. \\ &\left. \left. -3\mu_1^3 \left(13 J \cos (\Phi )+15 \sin (\Phi )\Delta \right)\right)\right] / \left[ \mu_1^{5/2} \left(225 \mu_1^6-104 J^2 \mu_1^3+12 J^4\right) \right].
\end{split}
\end{equation}
The coefficient $a_5(\Theta,\Phi,\Psi)$ is not fixed by the $R$ geodesic equation, but by the null condition. The coefficients $b_3^{(1)}$ and $\tilde{b}_3^{(1)}$ determined above are precisely those found by solving the geodesic equation for $\theta$ directly in section \ref{justify}.

We will not list the higher order coefficients, as they are somewhat long and unenlightening.

\subsection{Lack of smoothness} 
It is now possible to expand $K^2$ in $\lambda$ and thus determine the smoothness of the metric. The coefficients listed above are all that is necessary for this purpose. In fact, we only need $w_0, a_1, b_3^{(1)}$ and $\tilde{b}_3^{(1)}$. The norm of $K$ is
\begin{equation}
K^2 = -H(R,\theta,\phi)^{-2} \omega_{\psi}^2 + H(R,\theta,\phi) R^2 \cos^2 \theta
\end{equation}
which expands out to give
\begin{equation}
\label{K2}
K^2 = \cos^2 \Theta \mu_1 - \frac{J^2 \cos^4 \Theta}{4 \mu_1^2} + \frac{\cos^2 \Theta h_0 (J^2 \cos^2 \Theta + 2 \mu_1^3) \Delta }{2\mu_1^4}\lambda + \mathcal{O}(F(\ln \lambda) \lambda^{3/2}).
\end{equation}
As the terms above order $\lambda$ will contain factors of $\sin (m w_0 \ln \lambda)$ and $\cos (m w_0 \ln \lambda)$ (with $m$ some positive integer) we will use $\mathcal{O}(F(\ln \lambda) \lambda^{n/2})$ for a term that is morally of order $\lambda^{n/2}$ to remind ourselves of the $\ln \lambda$ dependence. We understand $F(\ln \lambda)$ to be a periodic function of $\ln \lambda$. The order $F(\ln \lambda) \lambda^{3/2}$ term in (\ref{K2}) is
\begin{equation}
\begin{split}
&\frac{J^2 \cos ^2(\Theta ) \sin (\Theta ) h_1 \Delta^{3/2} \left(5 J^2 \cos^2(\Theta )-(9 \cos (2 \Theta )+11) \mu_1^3\right)}{\mu_1^{9/2} \left(2J^2 - 9 \mu_1^3\right)} \sin \left( \Phi - \frac{J \ln \lambda}{2\Delta} \right) \lambda^{3/2} \\ &+ \frac{3 J \cos ^2(\Theta ) \sin (\Theta ) h_1 \left(J^2\cos ^2(\Theta )-2 \mu_1^3\right) \Delta^{5/2}}{\mu_1^{9/2} \left(9\mu_1^3 - 2J^2\right)} \cos \left( \Phi - \frac{J \ln \lambda}{2\Delta} \right) \lambda^{3/2}.
\end{split}
\end{equation}
Thus $K^2$ is not twice continuously differentiable. As argued earlier, lack of smoothness in the norm of this Killing vector field implies the metric is not smooth. In particular we now see that the metric is not $C^2$. This is worse than the result for an axial configuration in five dimensions where all the extremal black holes are static. In that case a similar argument involving the Killing fields on $S^2$'s in the geometry implies that the metric at the horizon is not $C^3$. We can see this immediately from the expression above simply by setting $J=0$: the order $\lambda^{3/2}$ term vanishes, and the next order term is an integer power of $\lambda$. As there is obviously no dependence on $\ln \lambda$ in the static case, the order $\lambda^2$ term is twice continuously differentiable. The order $\lambda^{5/2}$ term, however, does not vanish, and hence the metric is not $C^3$.

\subsection{Gaussian null metric}
We have seen that the metric at the horizon is not $C^2$. We will now show that it is $C^1$ there by transforming the metric to Gaussian null coordinates. Thinking of the functions in section \ref{seriesansatz} as coordinate transformations, it is straightforward to transform the metric from the polar coordinates $\{t,R,\theta,\phi,\psi\}$ to the Gaussian null system $\{v,\lambda,\Theta,\Phi,\Psi\}$. We have
\begin{equation}
dt = dv - d \left( \int (H^2 + \dot{\omega}) d\lambda \right)
\end{equation}
from the equation for $\dot{t}$. We can find $H$ as a function of our new coordinates $\lambda, \Theta, \Phi$ and $\Psi$ by substituting the expansions of the old coordinates, using the coefficients found before. Similarly we can take derivatives of $R(\lambda, \Theta, \Phi, \Psi)$ to find $dR$ in terms of the new coordinates, and we can do the same for $\theta$, $\phi$ and $\psi$. Thus we find a metric in the Gaussian null coordinate system. By construction, a metric in these coordinates must have $g_{\lambda \lambda} = 0$, $g_{v\lambda} = 1$ and $g_{\lambda x^i} = 0$ (for outgoing null geodesics), where the $x^i$ are coordinates on the $S^3$.

In practice we cannot take the expansions in section \ref{seriesansatz} to infinite order; we truncate the expansions at the order required to find a $C^1$ metric (as we now know the metric is not $C^2$). As we are truncating our series expansion ansatz, the metric will only take the Gaussian null form up to a certain order. We assume that inclusion of all higher order terms in the ansatz will allow us to find a truly Gaussian null form for the metric. For this reason we refer to the metric as being in ``nearly Gaussian null" form.

\subsubsection{Metric components}
We now give the form of the metric in our ``nearly Gaussian null" coordinate system:
\begin{equation}
\label{metriccomps}
\begin{split}
g_{vv} &= -\frac{\Delta^2}{\mu _1^4} \lambda^2 + \mathcal{O}(F(\ln \lambda) \lambda^{5/2}) \\
g_{v\lambda} &= 1 + \mathcal{O}(F(\ln \lambda) \lambda^{7/2}) \\
g_{v\Theta} &= \mathcal{O}(F(\ln \lambda) \lambda^{5/2}) \\
g_{v\Phi} &= \frac{-\sin^2 \Theta J \Delta}{2\mu_1^3} \lambda + \frac{J \sin^2 \Theta h_0 (14\mu_1^3 - 5 J^2)}{4\mu_1^5} \lambda^2 + \mathcal{O}(F(\ln \lambda) \lambda^{5/2}) \\
g_{v\Psi} &= \frac{-\cos^2 \Theta J \Delta}{2\mu_1^3} \lambda + \frac{J \cos^2 \Theta h_0 (14\mu_1^3 - 5 J^2)}{4\mu_1^5} \lambda^2 + \mathcal{O}(F(\ln \lambda) \lambda^{5/2}) \\
g_{\lambda\lambda} &= \mathcal{O}(\lambda^{3/2}) \\
g_{\lambda \Theta} &= \mathcal{O}(\lambda^{5/2}) \\
g_{\lambda \Phi} &= \mathcal{O}(\lambda^{5/2}) \\
g_{\lambda \Psi} &= \mathcal{O}(\lambda^{5/2}) \\
g_{\Theta \Theta} &= \mu_1 + \frac{h_0\Delta}{\mu_1} \lambda + \frac{2 J^2 \sin (\Theta ) h_1 \Delta^{3/2}}{\mu_1^{3/2} \left(9\mu_1^3 - 2J^2\right)} F_1(\Phi,\ln \lambda) \lambda^{3/2} + \frac{6 J \sin (\Theta ) h_1 \Delta^{5/2}}{\mu_1^{3/2} \left(2 J^2-9\mu_1^3\right)} F_2(\Phi,\ln \lambda) \lambda^{3/2} + \mathcal{O}(F(\ln \lambda) \lambda^2) \\
g_{\Theta \Phi} &= \frac{3 J \cos (\Theta ) \sin ^2(\Theta ) h_1 \sqrt{\Delta} \left(12 \mu_1^6-7 J^2 \mu_1^3+J^4\right)}{\mu_1^{9/2} \left(9 \mu_1^3 - 2J^2\right)} F_1(\Phi,\ln \lambda) \lambda^{3/2} \\ &+ \frac{3 J^2\cos (\Theta ) \sin ^2(\Theta ) h_1 \Delta^{7/2}}{\mu_1^{9/2} \left(9 \mu_1^3-2 J^2\right)} F_2(\Phi,\ln \lambda) \lambda^{3/2} + \mathcal{O}(F(\ln \lambda) \lambda^2)
\end{split}
\end{equation}
\begin{equation}
\begin{split}
g_{\Theta \Psi} &= \frac{3 J \cos ^3(\Theta ) h_1 \sqrt{\Delta} \left(12 \mu_1^6-7 J^2 \mu_1^3+J^4\right)}{\mu_1^{9/2} \left(9 \mu_1^3 - 2J^2\right)} F_1(\Phi,\ln \lambda) \lambda^{3/2} \\ &+ \frac{3 J^2 \cos ^3(\Theta ) h_1 \Delta^{7/2}}{\mu_1^{9/2} \left(9 \mu_1^3-2 J^2\right)} F_2(\Phi,\ln \lambda) \lambda^{3/2} + \mathcal{O}(F(\ln \lambda) \lambda^2)
\end{split}
\end{equation}
and
\begin{equation}
\begin{split}
g_{\Phi \Phi} &= \sin ^2(\Theta ) \mu_1-\frac{J^2 \sin ^4(\Theta )}{4 \mu_1^2} + \frac{\sin ^2(\Theta ) h_0 \left(2 \mu_1^3+J^2 \sin^2(\Theta )\right) \Delta}{2 \mu_1^4}\lambda \\ &+ \frac{J^2 \sin ^3(\Theta ) h_1 \Delta^{3/2} \left(J^2 (5 \cos (2 \Theta )+7)-2 (9 \cos (2 \Theta )+13) \mu_1^3\right)}{2 \mu_1^{9/2} \left(9 \mu_1^3 - 2J^2\right)} F_1(\Phi,\ln \lambda) \lambda^{3/2} \\ &+ \frac{3 J \sin ^3(\Theta ) h_1 \left(J^2 (\cos (2 \Theta )+3)-8 \mu_1^3\right) \Delta^{5/2}}{2 \mu_1^{9/2} \left(2 J^2-9\mu_1^3\right)} F_2(\Phi,\ln \lambda) \lambda^{3/2} + \mathcal{O}(F(\ln \lambda) \lambda^2) \\
g_{\Phi \Psi} &= -\frac{J^2 \cos ^2(\Theta ) \sin ^2(\Theta )}{4 \mu_1^2} + \frac{J^2 \cos ^2(\Theta ) \sin ^2(\Theta ) h_0 \Delta}{2 \mu_1^4} \lambda \\ &+ \frac{J^2 \cos ^2(\Theta ) \sin (\Theta ) h_1 \Delta^{3/2} \left(J^2 (5 \cos (2\Theta )+1)-6 (3 \cos (2 \Theta )+1) \mu_1^3\right)}{2 \mu_1^{9/2} \left(9 \mu_1^3 - 2J^2\right)} F_1(\Phi,\ln \lambda) \lambda^{3/2} \\ &+ \frac{3 J \cos ^2(\Theta ) \sin (\Theta ) h_1 \left(J^2\cos ^2(\Theta )-3 \mu_1^3\right) \Delta^{5/2}}{\mu_1^{9/2} \left(2 J^2-9\mu_1^3\right)} F_2(\Phi,\ln \lambda) \lambda^{3/2} + \mathcal{O}(F(\ln \lambda) \lambda^2)
\end{split}
\end{equation}
and finally
\begin{equation}
\begin{split}
g_{\Psi \Psi} &= \cos ^2(\Theta ) \mu_1-\frac{J^2 \cos ^4(\Theta )}{4 \mu_1^2} + \frac{\cos ^2(\Theta ) h_0 \left(2 \mu_1^3+J^2 \cos^2(\Theta )\right) \Delta}{2 \mu_1^4}\lambda \\ &+ \frac{J^2 \cos ^2(\Theta ) \sin (\Theta ) h_1 \Delta^{3/2} \left(5 J^2 \cos^2(\Theta )-(9 \cos (2 \Theta )+11) \mu_1^3\right)}{\mu_1^{9/2} \left(2J^2 - 9 \mu_1^3\right)} F_1(\Phi,\ln \lambda) \lambda^{3/2} \\ &+ \frac{3 J \cos ^2(\Theta ) \sin (\Theta ) h_1 \left(J^2\cos ^2(\Theta )-2 \mu_1^3\right) \Delta^{5/2}}{\mu_1^{9/2} \left(9\mu_1^3 - 2J^2\right)} F_2(\Phi,\ln \lambda) \lambda^{3/2} + \mathcal{O}(F(\ln \lambda) \lambda^2)
\end{split}
\end{equation} 
where
\begin{equation}
\begin{split}
F_1(\Phi,\ln \lambda) &= \sin \left(\Phi - \frac{J \ln \lambda}{2 \Delta}\right) \\
F_2(\Phi,\ln \lambda) &= \cos \left(\Phi - \frac{J \ln \lambda}{2 \Delta}\right).
\end{split}
\end{equation}
We see that terms of order $F(\ln \lambda) \lambda^{3/2}$ appear in all the angular components of the metric, thus the metric is $C^1$ at the horizon: only once differentiable. This is the main result of this paper. As stated earlier, the components $g_{\lambda \lambda}$, $g_{\lambda \Theta}$, $g_{\lambda \Phi}$ and $g_{\lambda \Psi}$ are vanishing in a true Gaussian null coordinate system. The presence of non-zero terms here is purely an artefact of truncating the series ansatz.

\subsection{Maxwell Field}
The Maxwell field potential is given by
\begin{equation}
A = \frac{\sqrt{3}}{2}H^{-1}(dt + \omega).
\end{equation}
In the Gaussian null coordinate system, $\{v, \lambda, \Theta, \Phi, \Psi\}$, this will obviously have the form
\begin{equation}
A = A_v dv + A_{\lambda} d\lambda + A_{\Theta} d\Theta + A_{\Phi} d\Phi + A_{\Psi} d\Psi.
\end{equation}
Performing the coordinate transformation using the expansions we have determined for $R, \theta, \phi$ and $\psi$, as well as (\ref{vdef}), we find
\begin{equation}
\begin{split}
A_v &= \frac{\sqrt{3} \Delta}{2 \mu_1^2} \lambda + \frac{3\sqrt{3} h_0 \left(J^2-2 \mu_1^3\right)}{4 \mu_1^4} \lambda^2 + \mathcal{O}(F(\ln \lambda) \lambda^{5/2})\\
A_{\lambda} &= -\frac{\sqrt{3} \mu_1^2}{2 \Delta} \lambda^{-1} +\frac{3 \sqrt{3} h_0 \left(J^2-2 \mu _1^3\right)}{4\Delta^2} + \frac{4 \sqrt{3} \sin (\Theta ) h_1 \left(J^2-5 \mu_1^3\right) \sqrt{\Delta}}{\sqrt{\mu_1} \left(25 \mu_1^3-6 J^2\right)} F_1(\Phi,\ln \lambda) \lambda^{1/2} \\ &+ \frac{4 \sqrt{3} J \sin (\Theta ) h_1 \Delta^{3/2}}{\sqrt{\mu_1} \left(25 \mu_1^3 - 6J^2\right)} F_2(\Phi,\ln \lambda) \lambda^{1/2} + \mathcal{O}(F(\ln \lambda) \lambda)\\
A_{\Theta} &= \mathcal{O}(F(\ln \lambda) \lambda^{3/2})\\
A_{\Phi} &= \mathcal{O}(F(\ln \lambda) \lambda^{3/2})\\
A_{\Psi} &= \frac{\sqrt{3} J \cos ^2(\Theta )}{4 \mu_1} -\frac{\sqrt{3} J \cos ^2(\Theta ) h_0 \Delta}{4 \mu_1^3} \lambda + \mathcal{O}(F(\ln \lambda) \lambda^{3/2}).
\end{split}
\end{equation}
The order $\lambda^{-1}$ and $\lambda^0$ terms in $A_{\lambda}$ are pure gauge, thus the Maxwell field strength is continuous at the horizon but not differentiable as $F_{\lambda \Theta} = \mathcal{O}(F(\ln \lambda) \lambda^{1/2}$. Typically we expect that for Einstein's equations to make sense they must be at least continuous. As the field strength is continuous at the horizon, the Einstein tensor must be continuous there, and Einstein's equations are satisfied.

\subsection{Parallel propagation of the Riemann tensor}
The fact that the metric is not $C^2$ implies the existence of a curvature singularity at the horizon. We will now show that there is a parallely propagated curvature singularity there. The leading order behaviour of a particular component of the Riemann tensor for small $\lambda$ is
\begin{equation}
R_{\lambda \Theta \lambda \Theta} = -\frac{1}{2}\partial_{\lambda }^2 g_{\Theta \Theta} + \ldots
\end{equation}
where the ellipsis denotes subleading terms. Referring back to (\ref{metriccomps}) we see that this will go as $\mathcal{O}(\lambda^{-1/2} \ln \lambda)$ and is therefore divergent at the horizon. As the metric is $C^1$, no other Riemann tensor components will be more divergent than this.

Now we construct an orthonormal basis on the horizon, which we can parallely propagate along the null geodesic with tangent vector $U = \partial / \partial \lambda$. First, we set $e^0 = d\lambda$ and $e^1 = dv$. Finally we take $e^2 = g_{\Theta \Theta}^{1/2} d\Theta$ so we have $e^0 \cdot e^1 = 1$, $e^0 \cdot e^2 = 0$, $(e^0)^2 = (e^1)^2 = 0$ and $(e^2)^2 = 1$ on the horizon. This frame is extended off the horizon by demanding that the basis one-forms are parallely propagated along the null geodesic: $U \cdot \nabla e_{1,2} = 0$, which preserves the orthogonality relations. Note that while the basis one-forms given above are valid only on the horizon, $e_0 = U$ everywhere.

In this basis then, we can find a particular component of the Riemann tensor
\begin{equation}
R_{0202} = e_0^{\;\; \lambda} e_2^{\;\; \Theta} e_0^{\;\; \lambda} e_2^{\;\; \Theta} R_{\lambda \Theta \lambda \Theta} + \ldots
\end{equation}
which gives
\begin{equation}
R_{0202} = g_{\Theta \Theta}^{-1} R_{\lambda \Theta \lambda \Theta} = -\frac{1}{2\mu_1} \partial_{\lambda}^2 g_{\Theta \Theta} + \ldots
\end{equation}
at leading order. This is clearly divergent as $\lambda \to 0$ and therefore there is a parallely propagated curvature singularity at the horizon.

\section{Higher Derivatives}
\label{fifth}
We now return to the static case of \cite{Candlish} in the context of higher derivative gravity theories. For this section we will very closely follow the work of \cite{higherderiv}, including their notation and metric signature. The details of higher derivative corrections to five dimensional supergravity resulting from compactification of string theory on Calabi-Yau manifolds may be found in that reference. All of the solutions we have discussed so far, including the extremal static black holes, are solutions to two-derivative five dimensional (minimal) supergravity. In \cite{higherderiv} the four-derivative correction to this action is discussed, and the consequences to the known single black hole solutions of the two-derivative theory are investigated in detail. Here we will present a simple extension of that work to consider the effects of this correction on the smoothness of axial multi-static black hole solutions.

\subsection{Very special geometry}
To begin with we briefly summarise some technical results and conventions. The details of the study of Calabi-Yau moduli spaces may be found in many references (see \cite{higherderiv} and references therein).

Five dimensional supergravity results from a compactification of M-theory on a Calabi-Yau 3-fold $CY_3$. First, let $J_I$ provide a basis of closed $(1,1)$ forms spanning the Dolbeault cohomology group $H^{(1,1)}(CY_3)$ of dimension $h^{(1,1)}$. The K\"ahler form $J$ on $CY_3$ may then be expanded as
\begin{equation}
J = M^I J_I, \quad I = 1 \ldots h^{(1,1)}.
\end{equation}
The coefficients of this expansion $M^I$ are known as \emph{K\"ahler moduli} and may be thought of as giving the volumes of a set of two-cycles in the Calabi-Yau. These two-cycles are the homology duals of the $(1,1)$ forms $J_I$. The \emph{intersection numbers} of the Calabi-Yau are given by
\begin{equation}
c_{IJK} = \int_{CY_3} J_I \wedge J_J \wedge J_K.
\end{equation}
This quantity may be thought of as counting the number of triple intersections of a set of four-cycles which are dual to the two-cycles discussed previously. Further to this we can write
\begin{equation}
\label{mdown}
M_I = \frac{1}{2} \int_{CY_3} J \wedge J \wedge J_I = \frac{1}{2}c_{IJK} M^J M^K
\end{equation}
which may be thought of as a dual moduli giving the volume of the $I$-th four-cycle.

It turns out that after compactification we have a set of gauge fields $A^I$ with field strengths $F^I$. One combination of the moduli lies in a supersymmetry hypermultiplet. This combination is in fact the total volume of the Calabi-Yau. As the hypermultiplets are decoupled from the other fields we can simply choose the value of this volume, giving us the \emph{very special geometry} constraint
\begin{equation}
\label{vsg}
\mathcal{N} = \frac{1}{6} c_{IJK} M^I M^J M^K = 1.
\end{equation}
Including higher derivative terms in the action leads to corrections to this constraint.

\subsection{Two-derivative solutions}
As stated in the introduction, the supersymmetric solutions of five dimensional supergravity, coupled to an arbitrary number of vector fields, fall into two classes, depending on whether a Killing vector constructed from the Killing spinor is timelike or null. We are interested in the timelike case, as the black hole solutions we have been discussing fall into this class. Following the notation of \cite{higherderiv}, the general solution for the timelike case is\footnote{To follow \cite{higherderiv} we have switched to a metric of negative signature.}
\begin{equation}
\begin{split}
ds^2 = e^{4U(x)}(dt + \omega)^2 - e^{-2U(x)}ds^2_B, \\
F^I = d[M^I e^{2U} (dt + \omega)] + G^I.
\end{split}
\end{equation}
where $ds^2_B$ denotes the line element on a four dimensional ``base space" which must be hyperK\"ahler due to supersymmetry. The function $U$, the one-form $\omega$ and the two-forms $G^I$ are all defined on this base space. If the base space is taken to be of Gibbons-Hawking type, this allows us to write the supersymmetric solutions in terms of a set of harmonic functions $(H^0, H^I; H_0, H_I)$ on $\mathbb{R}^3$. The equations from supersymmetry (the BPS equations) then give relations between these harmonic functions, $U$, $\omega$ and $G^I$.

\subsubsection{Static black holes}
Static five dimensional black holes are given by
\begin{equation}
H^0 = \frac{1}{|\vec{x}|}, \quad H^I = 0; \quad H_0 = 0, \quad H_I = H_I^{\infty} + \frac{q_I}{4|\vec{x}|}
\end{equation}
and $\omega = G^I = 0$. The harmonic function $H^0$ tells us that the base space is just $\mathbb{R}^4$ in this case, and the functions $H_I$ give the locations of the charge centres carrying the charges $q_I$, associated to the gauge fields $A^I$. $H_I^{\infty}$ gives the asymptotic value of the moduli $M_I$. The only non-trivial BPS equation in this case (with a two-derivative action) is
\begin{equation}
\label{mup}
M_I e^{-2U} = H_I.
\end{equation}
So, for a given compactification, we can find the moduli and the metric function $U$ in terms of the harmonic functions $H_I$. We again follow the example in \cite{higherderiv}: that of a compactification on $T^2 \times K3$. In this case the triple intersection numbers are $c_{1ij} = c_{ij}$ with $i,j = 2,\ldots,23$. Inverting the relation (\ref{mdown}) gives
\begin{equation}
\label{invert}
M^1 = \sqrt{\frac{c^{ij}M_iM_j}{2M_1}}, \quad M^i = c^{ij} M_j \sqrt{\frac{2M_1}{c^{kl}M_kM_l}}.
\end{equation}
Substituting (\ref{mup}) into these expressions we find
\begin{equation}
M^1 = \left(\frac{e^{2U}c^{ij}H_iH_j}{2H_1}\right)^{1/2}, \quad M^i = \left( \frac{e^{2U}c^{ij}H_iH_j}{2H_1}\right)^{-1/2}e^{2U}c^{ij}H_j.
\end{equation}
Now we consider a 3 charge black hole with $q_1 = q_2 = q_3 = q$, so $H_1 = H_2 = H_3 = H$, with all other asymptotic moduli and charges vanishing, so $q_i = H_i = 0$ for $i = 4 \ldots 23$. Thus we have
\begin{equation}
M^1 = M^2 = M^3 = e^{U}H^{1/2}.
\end{equation}
The BPS equations get us this far, but to find $U$ we need another condition: the very special geometry constraint (\ref{vsg}). Substituting the moduli into this condition we get
\begin{equation}
M^1M^2M^3 = 1 = e^{3U}H^{3/2} \quad \Rightarrow \quad e^{-2U} = H.
\end{equation}
Therefore in this example the metric is given by
\begin{equation}
ds^2 = H^{-2}dt^2 - Hds^2(\mathbb{R}^4),
\end{equation}
which is simply the metric for an extremal static black hole solution to Einstein-Maxwell theory in five dimensions. Similarly we could consider the harmonic functions $H_1, H_2$ and $H_3$ to have multiple charge centres. If each of the $i$-th charge centres has all three charges located there (and they are all equal), we would find the multi-centre static black hole solutions considered in \cite{Candlish} (albeit with a different metric signature).

\subsection{Four-derivative correction}
The four-derivative term discussed in \cite{higherderiv} has an overall factor of $c_{2,I}$, the $I$-th component of the second Chern class of the Calabi-Yau, in the $J_I$ basis. For a compactification on $T^2 \times K3$ the components are $c_{2,i} = 0, c_{2,1} = 24$.

The higher derivative term modifies the non-trivial BPS equation (for a static solution) to
\begin{equation}
\label{BPScorr}
M_Ie^{-2U} - \frac{c_{2I}}{8}(\nabla U)^2 = H_I
\end{equation}
while the very special geometry constraint becomes
\begin{equation}
\label{vsgcorr}
\mathcal{N} - 1 + \frac{c_{2I}}{24}e^{2U} \left[ M^I \left(\nabla^2 U - 4(\nabla U)^2 \right) + \nabla_i M^I \nabla^i U \right] = 0,
\end{equation}
with $\nabla_i$ denoting the covariant derivative on the base space. As the base space is simply $\mathbb{R}^4$, we will use the polar coordinate system where
\begin{equation}
ds^2(\mathbb{R}^4) = dr^2 + r^2 (d\theta^2 + \sin^2 \theta (d\phi^2 + \sin^2 \phi d\psi^2))
\end{equation}
so the Laplacian is
\begin{equation}
\nabla^2 U = \frac{1}{r^3} \partial_r (r^3 \partial_r U(r,\theta)) + \frac{1}{r^2 \sin^2 \theta} \partial_{\theta} (\sin^2 \theta U(r,\theta))
\end{equation}
and
\begin{equation}
(\nabla U)^2 = (\partial_r U(r,\theta))^2 + \frac{1}{r^2} (\partial_{\theta} U(r,\theta))^2.
\end{equation}
In anticipation of using an axially symmetric multi-centre ansatz for $H$, we have included a $\theta$-dependence in the function $U$.

\subsubsection{Multi-centre higher derivative solution}
We will take the harmonic functions $H_1 = H_2 = H_3 = H$ to be
\begin{equation}
H = \frac{\mu_1}{r^2} + \sum_{n=0}^{\infty} h_n r^n Y_n(\cos \theta).
\end{equation}
This is simply (\ref{multiharmgegen}) in a different polar coordinate system. As before the spherical harmonics $Y_n(\cos \theta)$ are given by the Gegenbauer polynomials:
\begin{equation}
Y_n (\cos \theta) = C_n^1 (\cos \theta).
\end{equation}
This is precisely the multi-centre ansatz considered in \cite{Candlish}, which preserves an $SO(3)$ isometry group. Given this choice for the harmonic functions we can use (\ref{invert}), (\ref{BPScorr}) and (\ref{vsgcorr}) to find the function $U$. For the smoothness analysis, we only require $U$ as a series expansion for small $r$, thus we consider
\begin{equation}
U(r,\theta) = \ln (u_0) + \ln (r) + \sum_{i=1}^{\infty} u_i(\theta)r^i.
\end{equation}
It turns out that we only need coefficients up to, and including, $u_5(\theta)$ to determine the smoothness of the metric.

The boundary condition for the $r$ coordinate is that the leading order behaviour of $e^{-2U}$ goes like $r^{-2}$. We have already taken this into account in choosing the coefficient of the $\ln (r)$ term in $U$ to be unity. Regularity on the $S^3$ in the geometry gives the boundary conditions for the $\theta$ coordinate, i.e. we demand that the coefficients in the expansion of $U$ are regular at $\theta = 0, \pi$. In this way we solve the very special geometry constraint for the coefficients of $U$. They are found to be
\begin{equation}
\begin{split}
u_0 &= \frac{(\mu_1+3)^{1/6}}{\mu_1^{1/3}(\mu_1^2+2\mu_1+1)^{1/6}} \\
u_1 &= 0 \\
u_2 &= -\frac{\left(h_0+1\right) \left(\mu_1+2\right)}{2\mu_1\left(\mu_1+1\right)} \\
u_3 &= -\frac{\cos (\theta ) h_1 \left(3\mu_1^2+16 \mu_1+24\right)}{3\mu_1^2 \left(\mu_1+4\right)} \\
u_4 &= \frac{3 \mu_1^5+47 \mu_1^4+323 \mu_1^3+807 \mu_1^2+738 \mu_1+162}{4 \mu_1^2 \left(\mu_1+1\right){}^2 \left(3 \mu_1^3+29 \mu _1^2+69 \mu _1+27\right)} \\ &+ \frac{\mu_1 \left(\mu_1 \left(\mu_1 \left(\mu_1 \left(3 \mu_1+47\right)+323\right)+807\right)+738\right)+162}{4 \mu_1^2\left(\mu_1+1\right){}^2 \left(\mu_1+3\right) \left(\mu_1\left(3 \mu _1+20\right)+9\right)} h_0(1+\frac{1}{2}h_0) - \frac{\csc (\theta ) \sin (3 \theta ) \left(3 \mu_1^2+17 \mu_1+30\right)}{6 \mu_1 \left(\mu_1^2+4 \mu_1-5\right)} h_2 \\
u_5 &= -\frac{2 \cos (\theta ) \cos (2 \theta ) \left(\mu_1 \left(\mu_1+6\right)+12\right)}{\left(\mu_1-2\right) \mu_1 \left(\mu_1+6\right)} h_3 + \frac{\cos (\theta ) \left(h_0+1\right) \left(\mu_1 \left(\mu_1 \left(3 \mu_1 \left(\mu_1+18\right)+448\right)+1192\right)+960\right)}{3 \mu_1^3\left(\mu_1+1\right) \left(\mu_1+4\right) \left(\mu_1+8\right)} h_1
\end{split}
\end{equation}
which, given that
\begin{equation}
\label{met}
ds^2 = e^{4U}dt^2 - e^{-2U}(dr^2 + r^2 (d\theta^2 + \sin^2 \theta d\Omega_2^2 )),
\end{equation}
means we have determined the metric (for small $r$) for the axially symmetric multi-centre static extremal black hole solution in the presence of a four-derivative correction to the action.

\subsubsection{Smoothness of higher derivative solution}
Following the same procedure as discussed in \cite{Candlish} we solve the geodesic equations for the coordinates $r$ and $\theta$ as series expansions in $\lambda$. This allows us to investigate the smoothness of the area of an invariant $S^2$ in the geometry. The metric on such an $S^2$ is found by restricting the full metric to the space spanned by the $SO(3)$ Killing fields. As stated before, the Killing fields have the same differentiability as the full metric, so the metric on the $S^2$'s is as differentiable as the full metric. First we find that the coordinate expansions in terms of the affine parameter $\lambda$ along a null geodesic are
\begin{equation}
\begin{split}
r(\lambda) &= \frac{\sqrt{2} \mu_1^{1/6} \left(\mu_1+1\right)^{1/6}}{(\mu_1+3)^{1/12}} \lambda^{1/2} + \frac{\left(h_0+1\right)\left(\mu_1+2\right)}{2 \sqrt{2\mu_1} \left(\mu_1+1\right)^{1/2} (\mu_1+3)^{1/4}} \lambda^{3/2} \\ &+ \frac{4 \cos (\Theta ) h_1 \left(\mu_1+1\right)^{2/3}\left(3 \mu_1^2+16 \mu_1+24\right)}{15 \mu_1^{4/3}\left(\mu_1+3\right)^{1/3} \left(\mu_1+4\right)} \lambda^2 + \ldots \\
\theta(\lambda) &= \Theta -\frac{2 \sqrt{2} \sin (\Theta ) h_1 \left(\mu_1+1\right)^{1/2} \left(3 \mu_1^2+16 \mu_1+24\right)}{3 \mu_1^{3/2} \left(\mu_1+3\right)^{1/4} \left(\mu_1+4\right)} \lambda^{3/2} -\frac{\sin (2 \Theta ) h_2 \left(\mu_1+1\right)^{2/3}\left(3 \mu_1^2+17 \mu_1+30\right)}{\mu_1^{1/3}\left(\mu_1+3\right)^{1/3} \left(\mu_1^2+4 \mu_1-5\right)} \lambda^2 \\ &-\frac{8 \sqrt{2} (3 \cos (2 \Theta )+2) \sin (\Theta )\left(\mu_1+1\right)^{10/12} \left(\mu_1^2+6 \mu_1+12\right)}{5 \mu_1^{1/6} \left(\mu_1+3\right)^{5/12} \left(\mu_1^2+4 \mu_1-12\right)} h_3 \lambda^{5/2} \\ &+ \frac{\sin (\Theta ) \left(h_0+1\right) \left(\mu_1+1\right)^{10/12} \left(51 \mu_1^4+846 \mu_1^3+5672 \mu_1^2+14000 \mu_1+11136\right)}{15 \sqrt{2} \mu_1^{13/6} \left(\mu_1+3\right)^{5/12} \left(\mu_1^3+13\mu _1^2+44 \mu _1+32\right)} h_1 \lambda^{5/2} + \ldots.
\end{split}
\end{equation}
The coefficient of the $\mathcal{O}(\lambda)$ term in $\theta(\lambda)$ has been chosen to vanish thus implementing the initial condition $\dot{\theta} = 0$ on the geodesic at the horizon. The area of the invariant $S^2$ in the geometry,
\begin{equation}
A_2 = -e^{-2U} r^2 \sin^2 \theta,
\end{equation}
expands out to give
\begin{multline}
A_2 = -\frac{\sin ^2(\Theta ) \mu _1^{2/3} \left(\mu_1+1\right)^{2/3}}{\left(\mu_1+3\right)^{1/3}} - \frac{2 \sin ^2(\Theta ) \left(h_0+1\right) \left(\mu_1+2\right)}{\left(\mu_1+3\right)^{1/2}}\lambda \\ - \frac{\sin ^2(\Theta ) \mu_1^{1/3} \left(3 \mu_1^4+29 \mu_1^3-55 \mu_1^2-357 \mu_1-324\right)}{\left(\mu_1+1\right)^{2/3} \left(\mu_1+3\right){}^{2/3} \left(3 \mu_1^3+29 \mu_1^2+69 \mu_1+27\right)} (2h_0 + h_0^2 + 1) \lambda^2 \\ + \frac{4 \sin ^4(\Theta ) \mu_1^{1/3} \left(\mu_1+1\right)^{4/3} \left(3 \mu_1^2+17 \mu_1+30\right)}{3 \left(\mu_1+3\right){}^{2/3} \left(\mu_1^2+4 \mu _1-5\right)} h_2 \lambda^2 \\ + \frac{64 \sqrt{2} \cos (\Theta ) \sin ^4(\Theta ) \sqrt{\mu_1}\left(\mu_1+1\right)^{3/2} \left(\mu_1^2+6 \mu_1+12\right)}{5 \left(\mu_1+3\right){}^{3/4}\left(\mu_1^2+4 \mu_1-12\right)} h_3 \lambda^{5/2} \\ + \frac{256 \sqrt{2} \cos (\Theta ) \sin ^2(\Theta )\left(h_0+1\right) \left(\mu_1+1\right)^{1/2} \left(\mu_1^3+27 \mu_1^2+87\mu_1+72\right)}{15 \mu_1^{3/2}\left(\mu_1+3\right){}^{3/4} \left(\mu_1+4\right) \left(\mu_1+8\right)} h_1 (1+h_0) \lambda^{5/2} + \ldots
\end{multline}
which is clearly not $C^3$. Transforming to a Gaussian null coordinate system, we would expect to find a $C^2$ metric. Thus the central result of \cite{Candlish} is unchanged in the presence of the four-derivative term. This is to be expected as the higher derivative correction does not change the powers of $r$ in the function $U$ in such a way as to improve smoothness.

\section{Discussion}
In this paper we have demonstrated that the supersymmetric rotating five dimensional black hole solution known as BMPV does not exhibit a smooth event horizon in the presence of other extremal black holes. For the axial configuration discussed here the horizon has been demonstrated to be $C^1$ but not $C^2$. This is worse than the case for a set of static black holes, which has a $C^2$ (but not $C^3$) horizon.

Further to this we have demonstrated that the smoothness result for static black holes is not affected by the inclusion of a higher derivative term in the action. It is expected that this will also be true for the four-derivative correction to the BMPV solution.

The lack of smoothness present for higher dimensional black holes seems to be ubiquitous in situations where rotational symmetries of the single black hole solution are broken. In the work of this paper we have broken one of the $U(1)$ rotational isometries of the BMPV black hole. Interestingly, in the concentric black ring solution of \cite{Gauntlett}, a BMPV black hole may reside at the centre of all the rings. In this case both $U(1)$ isometries are preserved and the horizon is analytic.

As another example of non-analyticity arising from breaking of a rotational symmetry, consider the solution presented in \cite{bena}. In that case a supersymmetric black ring cohabits the spacetime with a BMPV black hole placed at the centre of the ring, but vertically displaced from it with respect to the plane of the ring. Again we clearly have a breaking of one of the $U(1)$ isometries of the BMPV black hole because of the presence of the black ring. Thus the metric in this case will depend on the coordinate $\phi$ which parameterises this direction. In \cite{bena} we indeed see that the rotation one form contains terms involving, for example, $\cos \phi$. This leads us to expect that the horizon of the BMPV black hole in this solution will be non-smooth, as we have already seen that the $\phi$ coordinate has a leading order dependence of $\ln \lambda$ along an ingoing null geodesic with affine parameter $\lambda$. Thus a Gaussian null coordinate system near the horizon of the BMPV black hole will contain $\cos (\ln \lambda)$ terms and therefore the metric will be non-analytic at $\lambda = 0$.

We can similarly argue that the horizon of the black ring in this solution will be non-smooth. The transformations of the angular coordinates required for a regular metric on the horizon of a single supersymmetric black ring are of the form
\begin{equation}
d\phi = d\phi' - \frac{C_0}{r}dr
\end{equation}
where $C_0$ is some constant. There is a similar transformation for the $\psi$ coordinate. Given this we immediately see that the $\phi$ coordinate will again depend on $\ln r$ close to the horizon of the black ring (located at $r=0$ in this coordinate system). As $r$ will depend on some power of the affine parameter $\lambda$ we have $\ln r \sim \ln \lambda$ and so again we expect to see terms such as $\cos (\ln \lambda)$ appearing for a Gaussian null metric covering the horizon of the black ring, and thus non-smoothness at some order in $\lambda$.

\subsubsection*{\centering Acknowledgements}
The author is supported by the University of Nottingham, and would like to thank H.S. Reall for suggesting the project and many helpful discussions. I am grateful to Masashi Kimura for pointing out the work in \cite{Kimura}. I would also like to thank DAMTP for their hospitality.

\end{document}